\documentclass{aa}
\usepackage{graphics}

\newcommand{\Vm}{V_{\mathrm{m}}}
\newcommand{\Sm}{S_{\mathrm{m}}}
\newcommand{\Em}{E_{\mathrm{m}}}
\newcommand{\vstar}{v_{\mathrm{*}}}
\newcommand{\pstar}{p_{\mathrm{*}}}

\newcommand{\jp}{j+\frac{1}{2}}
\newcommand{\jm}{j-\frac{1}{2}}

\def\spose#1{\hbox to 0pt{#1\hss}}
\def\gsim{\mathrel{\spose{\lower 3pt\hbox{$\mathchar"218$}}
          \raise 2.0pt\hbox{$\mathchar"13E$}}}
\def\lsim{\mathrel{\spose{\lower 3pt\hbox{$\mathchar"218$}}
          \raise 2.0pt\hbox{$\mathchar"13C$}}}

\begin{document}

\thesaurus{03 
          (  13.07.1 
           ; 02.08.1 
           ; 02.19.1 
           ; 02.18.8 
           ; 02.18.5 
          )}
\title{Gamma-ray bursts from internal shocks in a relativistic wind : an hydrodynamical study.}
\author{F. Daigne\thanks{Present address: Max--Planck--Institut for Astrophysics, Karl--Schwarzschild--Str. 1, D-85740 Garching bei M\"unchen, Germany} \and R. Mochkovitch}
\institute{Institut d'Astrophysique de Paris, CNRS, 98 bis Boulevard Arago, 75014 Paris, France}
\date{Received **.**.** / Accepted **.**.**}
\authorrunning{Daigne \& Mochkovitch}
\titlerunning{GRBs from internal shocks in a relativistic wind, an hydrodynamical study.}
\maketitle

\begin{abstract}
The internal shock model for gamma-ray bursts involves shocks taking place 
in a relativistic wind with a very inhomogeneous initial distribution of 
the Lorentz factor.
We have developed a 1D lagrangian hydrocode to follow the evolution of 
such a wind and the results we have obtained 
are compared to those of a simpler model presented in a recent paper 
(Daigne \& Mochkovitch \cite{Daigne2}) where all pressure waves are 
suppressed in the wind so that shells with different velocities only 
interact by direct collisions. The detailed hydrodynamical calculation
essentially confirms the conclusion of the simple model: 
the main temporal and spectral properties of gamma-ray bursts can be 
reproduced by internal shocks in a relativistic wind.\\  

\end{abstract}


\section{Introduction}
Since the discovery of the optical counterpart of GRB 970228 (van Paradijs et al. \cite{vanParadijs1}) the 
accurate localizations provided by the \textit{Beppo--SAX} satellite have led to the detection of the optical 
afterglow for more than ten gamma--ray bursts (hereafter GRBs).
The most spectacular result of these observations is to have provided a direct proof of the cosmological origin of GRBs. The detection of absorption lines at $z=0.835$ in the spectrum of GRB 970508 (Metzger et al. \cite{Metzger1}) followed by other redshift determinations (between $z=0.43$ and $z=3.41$) confirmed the indications 
which were already available from the \textit{BATSE} data showing a GRB distribution perfectly isotropic 
but non homogeneous in distance (Fishman and Meegan \cite{Fishman1} and references therein).

The energy release of GRBs with known redshifts extends from $E_{\gamma}= 2\ 10^{51} \frac{\Omega}{4 \pi}$ to 
$E_{\gamma} = 2\ 10^{54} \frac{\Omega}{4 \pi}\ \rm erg$. The solid angle $\Omega$ in which the emission is beamed is quite uncertain. 
A small $\Omega$ should reveal itself by a break after a few days in the afterglow light curve.  
A break is indeed observed 
in a few cases such as GRB 990510 (Harrison et al. \cite{Harrison1}) where $\frac{\Omega}{4 \pi}$ could be as small as 0.01. 
However, most afterglows do not show any break which means that $\Omega$ is usually not very small
($\frac{\Omega}{4 \pi}\sim 0.1$\,?).\\
The source of cosmic GRBs must therefore be able to release a huge energy in a very short time. Possible
candidates include the coalescence of two neutron stars (Eichler et al. \cite{Eichler1}; Paczy\'nski \cite{Paczynski1}), the disruption of the neutron star in a neutron star -- black hole binary (Narayan et al. \cite{Narayan1}; Mochkovitch et al. \cite{Mochkovitch1})  
or 
the collapse of a massive star (Woosley \cite{Woosley1}, Paczy\'nski \cite{Paczynski2}). 
In all these cases the resulting configuration is expected to be a stellar mass black hole surrounded 
by a thick disc. 
Since the power emitted by GRBs is orders of magnitude larger than the Eddington limit it cannot 
be radiated by a static photosphere. The released energy generates a fireball which then leads to the formation 
of a wind. Moreover, this wind has to become highly relativistic in order to avoid the compactness problem
and produce gamma--rays (Baring \cite{Baring1}; Sari \& Piran \cite{Sari1}). Values of the Lorentz factor 
as high as $\Gamma=100$--$1000$ are required, which limits the allowed amount of baryonic pollution to a remarkably low level. Only a few mechanisms have been proposed to produce a wind under such severe constraints : 
({\it i}) magnetically driven outflow originating from the disc or powered by the Blandford--Znajek (\cite{Blandford1}) process (Thomson \cite{Thomson1}; M\'esz\'aros \& Rees \cite{Meszaros2}; Daigne \& Mochkovitch \cite{Daigne3}; Lee et al. \cite{Lee1}) ; 
({\it ii}) reconnection of magnetic field lines in the disc corona (Narayan et al. \cite{Narayan1}) ; 
({\it iii}) neutrino--antineutrino annihilation in a funnel along the rotation axis of the system (M\'esz\'aros \& Rees \cite{Meszaros3}; Mochkovitch et al. \cite{Mochkovitch1}, \cite{Mochkovitch2}). 
Mechanisms ({\it i}) and ({\it ii}) require that the magnetic field in the disc reaches very high values 
$B \gsim 10^{15}\ \rm G$. Our preliminary study (Daigne \& Mochkovitch \cite{Daigne3}) of the wind emitted
from the disc shows 
that it can avoid baryonic pollution  
only if some very severe constraints on the dissipation in the disc and
the field geometry are satisfied. 
Some recent works (Ruffert et al. \cite{Ruffert1}) have also shown that mechanism ({\it iii}) 
is probably not efficient enough to power a gamma--ray burst, except may be for 
the shortest events.

When the wind has reached its terminal Lorentz factor, the energy is mainly stored in kinetic form and has to be converted back into gamma--rays. 
Two main ideas have been proposed to realize this conversion. 
The first one is the so-called external shock model (Rees \& M\'esz\'aros \cite{Rees1}; M\'esz\'aros \& Rees \cite{Meszaros4}). The wind is decelerated by the external medium, leading to a shock. Gamma--rays are emitted by the accelerated electrons in the shocked material through the synchrotron and/or inverse Compton mechanisms. This model has been studied in details (Fenimore et al. \cite{Fenimore1}; Panaitescu et al. \cite{Panaitescu1}; Panaitescu \& M\'esz\'aros \cite{Panaitescu2}) and seems unable to reproduce some important features of GRBs such as their strong temporal variability (see however Dermer \& Mitman \cite{Dermer1}). 
Conversely, the external shock model reproduces very well the delayed 
emission at lower energy from the afterglows (M\'esz\'aros \& Rees \cite{Meszaros5}; Wijers et al. 
\cite{Wijers1}). 

The second proposal is the internal shock model (Rees \& M\'esz\'aros \cite{Rees2}) where the wind is supposed to be formed initially with a very inhomogeneous distribution of the Lorentz factor. Rapid parts of the wind then catch up 
with slower ones leading to internal shocks where gamma--rays are 
again produced by synchrotron or inverse Compton radiation. We have started a 
study of 
this model in a previous paper (Daigne \& Mochkovitch \cite{Daigne2}, hereafter DM98) where the wind was simply made of a collection of ``solid'' shells 
interacting by direct collisions only (all pressure waves were suppressed). The very encouraging results we obtained had to be confirmed by a more detailed
study. We have therefore developed a relativistic hydrocode to follow
the evolution of the wind. We present the code and the main results in this
paper. We write in Sect. 2 the lagrangian equations of hydrodynamics in special relativity.
In Sect. 3 we describe the numerical method we use to solve them and 
we present the tests we performed to validate the method. 
We display our results in Sect. 4 and
Sect. 5 is the conclusion.


\section{Lagrangian equations of hydrodynamics in special relativity}
We write in a fixed frame the equations of mass, momentum and energy conservation in spherical symmetry :
\begin{equation}
\frac{\partial \Vm}{\partial t} - \frac{\partial\left(R^{2} v\right)}{\partial m} = 0\ ,
\label{eqVm}
\end{equation}
\begin{equation}
\frac{\partial \Sm}{\partial t} + R^{2} \frac{\partial P}{\partial m} = 0\ ,
\label{eqSm}
\end{equation}
\begin{equation}
\frac{\partial \Em}{\partial t} + \frac{\partial\left(R^{2} P v\right)}{\partial m} = 0\ ,
\label{eqEm}
\end{equation}
where $R$ and $t$ are respectively the spatial and temporal coordinates in the fixed frame.
The following quantities appear in Eqs. \ref{eqVm}--\ref{eqEm} : $P$ is the pressure in the fluid local rest frame, $v$ is the fluid velocity in the fixed frame and $\Vm$, $\Sm$ and $\Em$ are the specific volume, 
momentum density and energy density (including mass energy) in the fixed frame. These three quantities are related to quantities in the fluid local rest frame :
\begin{equation}
\Vm = \frac{1}{\rho \Gamma}\ ,
\end{equation}
\begin{equation}
\Sm = h \Gamma v\ ,
\end{equation}
\begin{equation}
\Em = h \Gamma - \frac{1}{\Gamma} \frac{P}{\rho}\ ,
\end{equation}
where $\Gamma = \frac{1}{\sqrt{1-v^{2}}}$ is the Lorentz factor, $\rho$ 
is the rest-mass density and $h = 1 + \epsilon + \frac{P}{\rho}$ is the 
specific enthalpy density ($\epsilon$ being the specific internal energy density). 
The lagrangian mass coordinate $m$ is defined by 
\begin{equation}
m = \int_{R_{\rm min}}^{R} \frac{R^{2}}{\Vm} dR\ ,
\end{equation}
$R_{\rm min}$ being the radius of the back edge of the wind.
The system of Eqs \ref{eqVm}--\ref{eqEm} is completed by the equation of state
\begin{equation}
P = \left(\gamma-1\right) \rho \epsilon\ ,
\end{equation}
the adiabatic index $\gamma$ being a constant.

In the non-relativistic limit ($v \to 0$ and $h \to 1$), the quantities 
$\Vm$, $\Sm$ and $\Em$ become equal to their newtonian counterparts 
$\frac{1}{\rho}$, $v$ and $E = 1+\epsilon+\frac{v^{2}}{2}$ and 
Eqs. \ref{eqVm}--\ref{eqEm} then reduce to the classical equations of 
lagrangian 
hydrodynamics in spherical symmetry. The great similarity between the 
relativistic and classical equations will allow us to use the powerful 
numerical methods which have been developed in classical hydrodynamics 
to follow the evolution of a fluid with shocks.    


\section{Numerical method}
\subsection{Extension of the PPM to 1D lagrangian relativistic hydrodynamics in spherical symmetry}
An extension of the Piecewise Parabolic Method of Colella and Woodward 
(\cite{Colella1}) to 1D eulerian relativistic hydrodynamics in planar 
symmetry has been already presented by Mart\'{\i} and M\"uller (\cite{Marti2}). 
We follow exactly the same procedure to extend the PPM to the lagrangian 
case in spherical symmetry.

We adopt $W = \left(\Vm , \Sm, \Em \right)$ as the set of variables. 
If the mass-averaged values $W_{j}^{n}$ of $W$ at time $t^{n}$ in each 
cell $\left[m_{\jm} , m_{\jp}\right]$ extending from $R_{\jm}^{n}$ 
to $R_{\jp}^{n}$ are known, the values $W_{j}^{n+1}$ at time 
$t^{n+1}=t^{n}+\Delta t$ are computed in four steps:

\noindent\underline{{\it a)} Reconstruction step}\\
The variables $U = \left(\frac{1}{\rho}, P, v\right)$ are obtained from $W$ in each cell by solving the following equation in $h$ : 
\begin{equation}
h^{2}+\left(\gamma-1\right)h-\gamma \Em \sqrt{\Sm^{2}+h^{2}} + \gamma \Sm^{2} = 0\ .
\end{equation}
Once $h$ is known, the other quantities are easily computed since 
$v = \frac{\Sm}{\sqrt{\Sm^{2}+h^{2}}}$ and 
$\frac{1}{\rho} = \Gamma \Vm$. These mass-averaged values are then 
interpolated by polynomials in the way described in the original paper by 
Colella and Woodward (\cite{Colella1}). We use the same modifications 
of the coefficients of the interpolation polynomials, leading to a steeper 
representation of discontinuities and a monotone representation of 
smoother parts.

\noindent\underline{{\it b)} Effective states}\\
At each interface $\jp$, two effective states $U_{\jp, L}$ and  $U_{\jp, R}$ ($L$ and $R$ denote the left and right sides of the interface) are constructed by mass-averaging these quantities in the region of cells $j$ and $j+1$ connected to the interface $\jp$ by a characteristic line during the time step.

\noindent\underline{{\it c)} Riemann solver}\\
At each interface $\jp$, the two effectives states $U_{\jp, L}$ and $U_{\jp, R}$ define a Riemann problem 
(two constant states separated by a discontinuity surface), which is known to give rise, like in newtonian 
hydrodynamics, to two new states $U_{\jp, L *}$ and $U_{\jp, R *}$ separated by a contact discontinuity and related to the initial states $U_{\jp, L}$ and $U_{\jp, R}$ either by a shock or a rarefaction wave. The common values of the pressure and the velocity of the two new intermediate states are given by the implicit equation
\begin{equation}
\vstar = v_{L*}(\pstar) = v_{R*} (\pstar)\ .
\end{equation}
An analytic expression (for a polytropic gas) of 
\begin{equation}
v_{S*}(p) = \left\lbrace
\begin{array}{cl}
\mathcal{R}_{S} (p) & {\rm if}\ p\le p_{S}\ \rm{(rarefaction\ wave)}\\
\mathcal{S}_{S} (p) & {\rm if}\ p\ge p_{S}\ \rm{(shock\ wave)}\\
\end{array}
\right.
\end{equation}
(where $S$ either refers to the $L$ or $R$ state) has been worked out by Mart\'{\i} 
and M\"uller (\cite{Marti1}). Equation (11) is solved using Brent's method 
to obtain the pressure $\pstar$ and the velocity $\vstar$ of the 
intermediate states at each interface.

\noindent\underline{{\it d)} Time advancement}\\
The quantities $W^{n}_{j}$ are calculated with numerical fluxes at each interface obtained from $\pstar$ and $\vstar$ in the same way than Colella and Woodward (\cite{Colella1}).
\subsection{Numerical tests}
\subsubsection{Relativistic shock tube}
\begin{figure}
  \resizebox{\hsize}{!}{\includegraphics{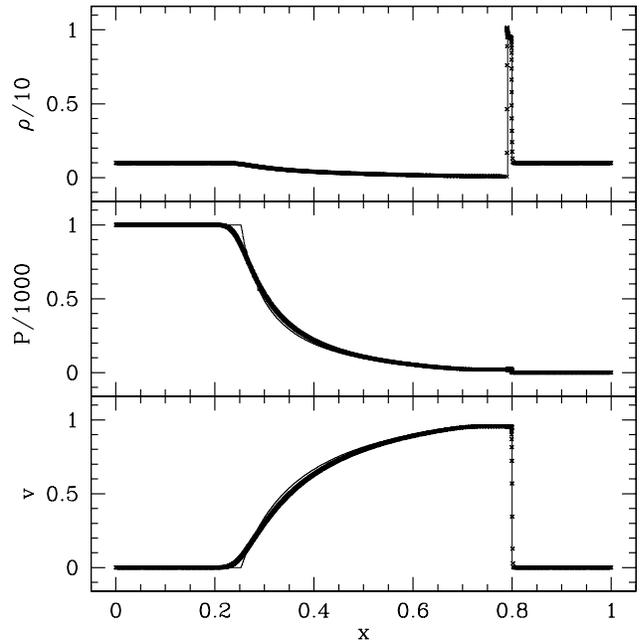}}
  \caption{\textit{Relativistic shock tube problem with 
$\rho_{L}=1$, $p_{L}=1000$ and $\rho_{R}=1$, $p_{R}=0.1$ : Exact (solid line) 
and numerical profiles of density, pressure and velocity at $t=0.303$. }}
  \label{FigTest1}
\end{figure}

We have sucessfully checked our code against two usual tests in 
relativistic hydrodynamics. The first one is the shock tube problem which 
is simply a Riemann problem in which the initial states are at rest. 
We present in Fig.\ref{FigTest1} the results for  
$\rho_{L}=1$, $p_{L}=1000$, $v_{L}=0$ and $\rho_{R}=1$, $p_{R}=0.1$, 
$v_{R}=0$. The adiabatic index is $\gamma=\frac{5}{3}$ and the discontinuity 
is initially located at $x=0.5$. The figure is plotted at a time $t=0.303$ 
for a grid of 1000 zones initially equally spaced.
The agreement between the exact and numerical profiles is  
satisfactory. The positions of the contact discontinuity and the shock 
are very accurate. The density, pressure and velocity of the post-shock 
state are also exact. However, the value of the density in the immediate 
vicinity of the contact discontinuity shows a small non-physical increase, 
which is more pronounced when the shock is stronger and disappears when the 
shock is weak (as in the shock tube problem with $\rho_{L}=10$, $p_{L}=13.3$ 
and $\rho_{R}=1$, $p_{R}=0$, which has been considered by several authors).
In the context of the internal shock model for GRBs, the shocks are only mildly 
relativistic and we do not 
observe any unexpected increase of the density in the results presented 
below.
\subsubsection{Spherical shock heating}
\begin{figure}
  \resizebox{\hsize}{!}{\includegraphics{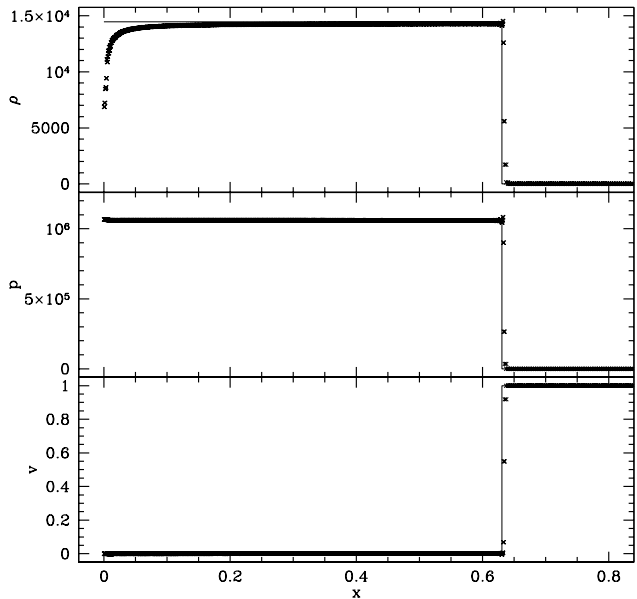}}
  \caption{\textit{Spherical shock heating : Exact (solid line) and numerical 
profiles of density, pressure and velocity at $t=1.90$. }}
  \label{FigTest2}
\end{figure}

This test consists in a cold fluid, which is initially homogeneous 
($\rho(R,0)=\rho_{0}$) and enters a sphere of radius 1 at constant velocity 
$v_{0}$. The fluid bounces at $R=0$ and is heated up. We present in 
Fig.\ref{FigTest2} the results for $\rho_{0} = 1$, $p_{0} = 10^{-6}$ and 
$v_{0} = -0.99999$ at $t=1.90$. The adiabatic index is $\gamma=\frac{4}{3}$ 
and we used a grid of 1000 zones initially equally spaced. In the considered 
case of a cold homogeneous fluid, an analytical solution is known. The shocked 
state is at rest with a density 
$\rho'_{0} = \frac{1}{\Gamma_{0}^{2}}\left(\frac{\gamma \Gamma_{0}+1}{\gamma-1}
\right)^{3} \rho_{0}$ and a pressure 
$p'_{0}=(\gamma-1)(\Gamma_{0}-1)\rho'_{0}$. At time $t$ the unshocked cold 
fluid of velocity $v_{0}$ has a distribution of density 
$\rho(R,t)=\rho_{0} \left(1+\frac{|v_{0}| t}{R}\right)^{2}$. 
The numerical profiles appear accurate except in the vicinity of the origin. 
The shock propagates with
 the correct velocity and the post-shock values of density and  
pressure are well reproduced. We therefore conclude that the treatment of the 
geometrical terms in Eqs \ref{eqVm}--\ref{eqEm} is correct. 
We have not tried to improve the computation near the center, which 
is not of major importance in the context of the internal shock model 
for GRBs where most of the emission takes place far from the 
origin.


\section{Results and discussion}
We have used our code to follow the evolution of a relativistic wind with 
a very inhomogeneous initial distribution of the Lorentz factor. The first 
results have been already presented for small values of the Lorentz factor 
$\Gamma \sim 40$ (Daigne \& Mochkovitch \cite{Daigne1}). Here we describe 
our results for the large Lorentz factors ($\Gamma \ge 100$) 
which are relevant for the study of GRBs. 
We first consider the case of a simple 
single-pulse burst.
\subsection{Initial state}
\begin{figure}[t]
  \resizebox{\hsize}{!}{\includegraphics{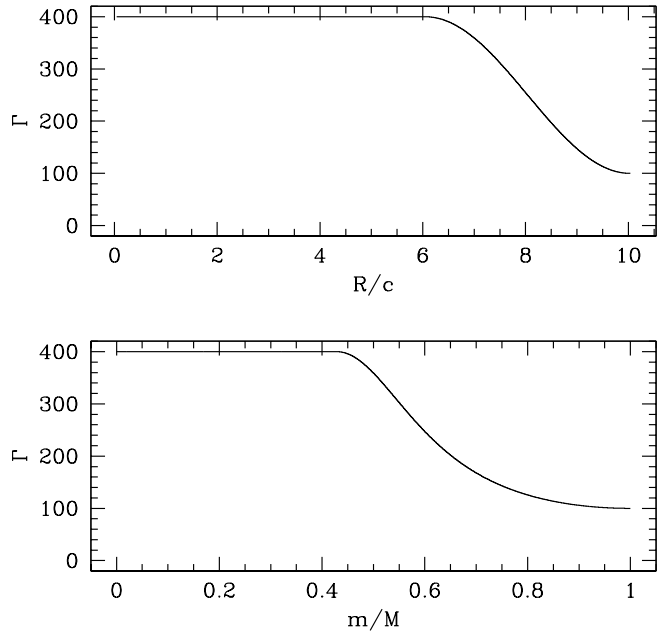}}
  \caption{\textit{Initial state at $t=10\ \rm s$ for $t_{W} = 10\ \rm s$. An energy $E= 2\ 10^{52} / 4 \pi\ \rm erg/sr$ has been injected into the wind, whose mass is $7.76\ 10^{28}\ \rm g$ (which corresponds to an 
average Lorentz factor $\bar{\Gamma} \sim 290$). 
The masses of the fast ($\Gamma = 400$) and "slow" parts ($\Gamma = 100 \to 400$) are equal. Upper panel : eulerian distribution of the Lorentz factor in the wind. Lower panel : corresponding lagrangian distribution.}}
  \label{FigInitial}
\end{figure}
Whatever the initial event leading to a GRB may be (NS--NS or NS--BH merger, 
``hypernova'', etc), the system at the end of this preliminary stage is 
probably made of a stellar mass black hole surrounded by a thick disc 
(the ``debris'' torus). We consider that $E$, a substantial fraction of the available energy of the system, is injected at a typical radius $R_{0}$ into a wind emitted during a duration $t_{W}$ with a mass flow $\dot{M}$. We do not discuss here the physical processes controlling $\dot{M}$, $t_{W}$ and $E$ but we assume that the baryonic load $\frac{1}{\eta} = \frac{\dot{M} t_{W} c^{2}}{E}$ is very small. The wind converts its internal energy into kinetic energy during its free expansion in the vacuum (the effect of the interstellar medium is negligible at this early stage) and accelerates until it reaches a Lorentz factor 
$\Gamma\simeq \eta$ at a typical radius $\Gamma R_{0}$ (M\'esz\'aros et al. \cite{Meszaros1}). This is where our simulation starts.

More precisely, we define our initial state as follows. We consider that from $t=0$ to $t=t_{W}$, 
a wind with a distribution of the Lorentz factor defined by 
\begin{equation}
\Gamma(t) = \left\lbrace\begin{array}{cl}
250-150 \cos{\left(\pi \frac{t}{0.4 t_{W}}\right)} & \rm if\ t \le 0.4\ t_{W}\\
400 & \rm if\ t \ge 0.4\ t_{W}\\ 
\end{array}\right.
\label{EqGammaInitial}
\end{equation}
has been produced by the source and that its back edge has reached $R_{\rm min}=400\,R_0=
1.2\ 10^4$ km (we adopt $R_{0} = 30\ \rm km$). 
We suppose that energy is 
injected at a constant rate $\dot{E}$ (we adopt 
$\dot{E}=\frac{2\ 10^{51}}{4 \pi}\ \rm erg.s^{-1}/sr$ in the following), so that the 
total energy injected into the wind simply equals $E = \dot{E}\ t_{W}$ and 
the injected mass flux is $\dot{M}(t) = \frac{\dot{E}}{\Gamma(t)\ c^{2}}$. 
The density profile in this 
initial state can be calculated if we assume that the internal energy is very small compared 
to the kinetic energy, which is indeed the case when the wind has reached its terminal 
Lorentz factor ($\frac{P}{\rho c^{2}} \ll 1$). The eulerian and lagrangian 
profiles of $\Gamma$ in the wind at $t=t_{W}$ are shown 
in Fig.\ref{FigInitial} for $t_{W} = 10\ \rm s$ and $E = 
\frac{2\ 10^{52}}{4 \pi}\ \rm erg/sr$. We have adopted 
$\frac{P}{\rho c^{2}}=10^{-3}$ and have checked that the results do not 
depend on this small value.
\begin{figure*}
\resizebox{9cm}{!}{\includegraphics{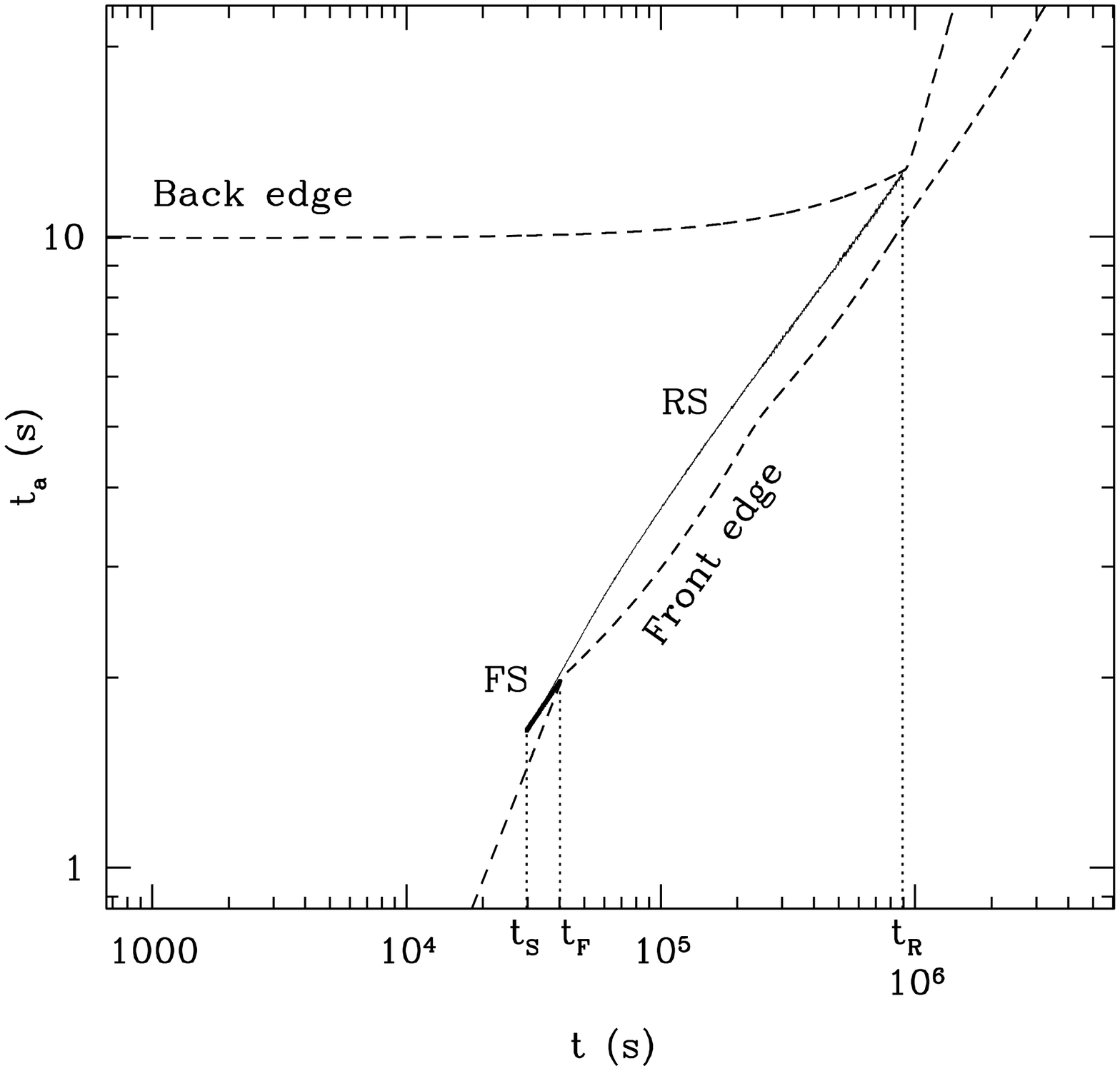}}
\resizebox{9cm}{!}{\includegraphics{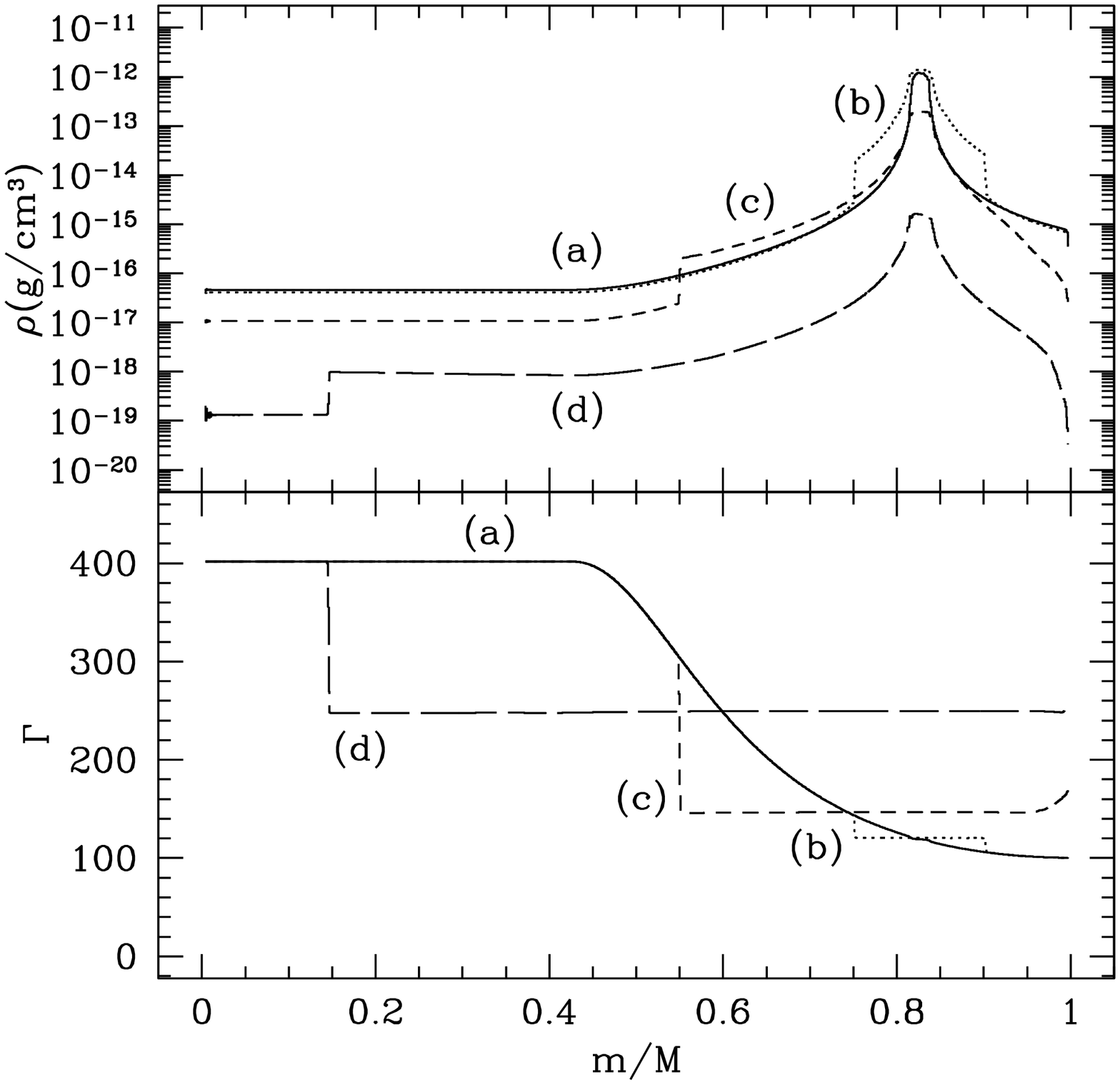}}
\caption{\textit{Left panel : } Paths of the back and front edges and of the 
forward and reverse shocks in the $t$--$t_{a}$ plane. The two shocks appear 
at $t_{S}$, the forward shock reaches the front edge very soon at $t_{F}$ 
and the reverse shock reaches the back edge later at $t_{R}$. 
\textit{Right panel : } Distribution of density $\rho$ and Lorentz factor 
$\Gamma$ at different times: (a) $t=2.5\ 10^{4}\ s$, just before the formation 
of the two shocks; (b) $t=3.0\ 10^{4}\ s$: the two shocks are clearly
visible; (c) $t=3.1\ 10^{4}\ s$ : the forward shock has just reached the 
front edge. The reverse shock has still more than one half of the mass 
to sweep; (d) $t=5.6\ 10^{5}\ s$: just before the reverse shock reaches the 
back edge.}
\label{FigDynamique}
\end{figure*}

\subsection{Dynamical evolution}
The fast part of the wind catches up with the slower one. The matter is 
strongly compressed is the collision region, the velocity gradient becomes very steep 
and at $t_{S} \sim 3\ 10^{3}\ t_{W}$, two shocks appear: a forward shock 
reaching the front edge at $t_{F} \sim 4\ 10^{3} t_{W}$ and a reverse shock 
reaching the back edge at $t_{R} \sim 9\ 10^{4} t_{W}$. The hot and dense 
matter behind these two internal shocks radiates and 
produces the observed burst. The radiation losses are not taken
into account in the dynamics, which is probably not a too severe approximation since the  
dissipated energy represents about  
10\% of the total kinetic energy of the wind.

When the two shocks have reached the edges, the evolution becomes unimportant 
regarding the emission of gamma-rays: two rarefaction waves 
develop at each edge and 
the wind continues to expand and cool. In fact, at this stage, the interstellar medium 
should absolutely be included in the calculation. An external shock 
propagates into the ISM which produces the afterglow and a reverse shock 
crosses the wind which can also lead to an observable emission. 
All these effects are not included in the present simulation, which is stopped
when the two internal shocks have reached the wind edges.

We present in Fig.\ref{FigDynamique} (left panel) the paths in a  
$t$ -- $t_{a}$ plot ($t_{a} = t - \frac{R}{c}$ is the arrival time of photons 
emitted at time $t_{e} = t$ on the line of sight at a distance $R$ from the source) 
of the two shocks 
and of the two edges of the wind. 
In the right panel the corresponding distributions of $\Gamma$ and $\rho$ 
are plotted at different times. 
\subsection{Gamma-ray emission and properties of the observed burst}
\subsubsection{Method of calculation}
\begin{figure}[t]
  \resizebox{\hsize}{!}{\includegraphics{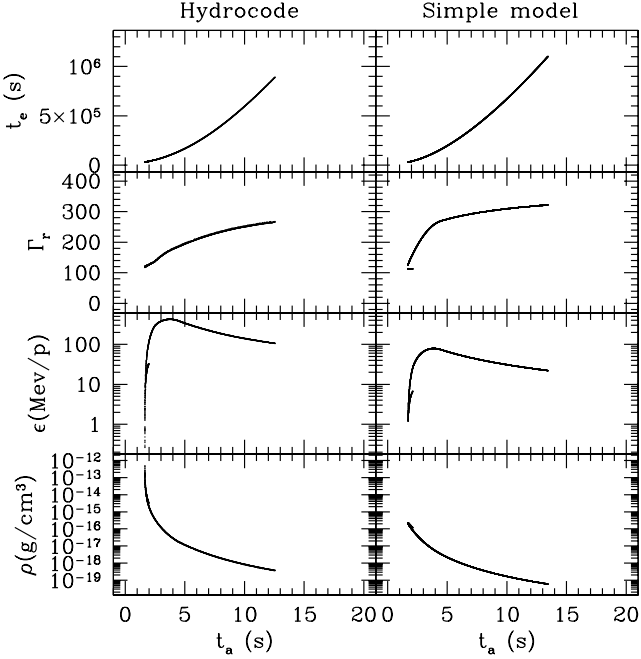}}
\caption{\textit{Single pulse burst ($10 s$) :} Emission time $t_{e}$, 
Lorentz factor of the emitting material $\Gamma_{r}$, dissipated energy 
per proton $\epsilon_{diss}$ and density of the shocked material $\rho$ 
as a function of arrival time $t_{a}$. Both contributions of the 
forward and reverse shocks are represented (the contribution of the 
forward shock is hardly visible).}
\label{FigQuantitesHydro}
\end{figure}
\begin{figure}[t]
\resizebox{\hsize}{!}{\includegraphics{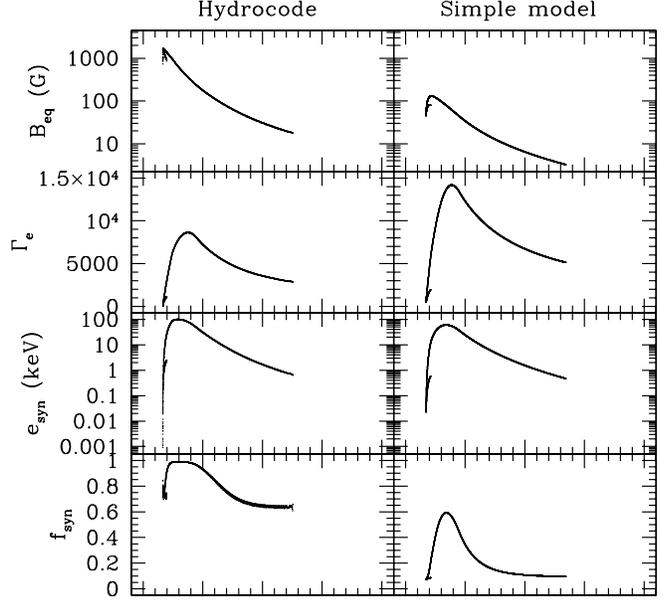}}
\caption{\textit{Single pulse burst ($10 s$) :} Magnetic field $B_{eq}$, Lorentz factor of the accelerated electrons $\Gamma_{e}$, synchrotron energy $e_{syn}$ and fraction of the energy which is radiated by the synchrotron process $f_{syn}$ as a 
function of arrival time $t_{a}$. Both contributions of the forward and reverse shocks are represented (the contribution of the forward shock is hardly visible).}
\label{FigEquipartition}
\end{figure}
Consider an internal shock located at a distance $R = R_{e}$ from the
source at a time $t=t_{e}$ in the fixed frame. 
The density $\rho_{*,S}$, the Lorentz factor $\Gamma_{*,S}$ and the specific 
internal energy $\epsilon_{*,S}$ of the shocked ($*$) and unshocked ($S$) 
material are known from our hydrodynamical simulation. This 
shock will produce a contribution to the GRB which will be observed at an
arrival time 
\begin{equation} 
t_{a}=t_{e}-\frac{R_{e}}{c}
\end{equation}
and which will last  
\begin{equation} 
\Delta t_{a}=\frac{R_{e}}{2c \Gamma_{r}^{2}}
\end{equation}
where $\Gamma_{r}$ is the
Lorentz factor of the emitting material for which we adopt
$\Gamma_{r}=\Gamma_{*}$. The luminosity of the shock is estimated 
by 
\begin{equation}
L_{sh} = \dot{M}_{sh}\ \Gamma_{*} \left(\epsilon_{*}-\epsilon_{S}\right)
\end{equation}
where $\dot{M}_{sh}$ is the mass flux across the shock and
$\epsilon_{diss}=\epsilon_{*}-\epsilon_{S}$ is 
the dissipated energy per unit mass in the frame of 
the shocked material.

Our code detects all the internal shocks present in the wind at a given time 
and saves their parameters in order to sum all the contributions to the 
emission and produce a synthetic gamma-ray burst. In a recent paper (DM98) 
we presented a simple model where the wind was idealized by a collection of 
``solid'' 
shells interacting by direct collision only (i.e. all pressure 
waves were neglected). We detailed in this previous paper our assumptions to treat the emission of a given shock. We adopt here the same assumptions. The magnetic field in the shocked material is supposed to reach equipartition values
\begin{equation}
B_{eq} = \sqrt{\alpha_{B}\ 8 \pi \rho \epsilon_{diss}}\ .
\end{equation} 
with $\alpha_{B}=\frac{1}{3}$. The Lorentz factor of the accelerated electrons is calculated using the expression given by Bykov \& M\'esz\'aros (\cite{Bykov1}) who consider the scattering of electrons by turbulent magnetic field fluctuations :
\begin{equation}
\Gamma_{e} = \left[ \frac{\alpha_{M}}{\zeta} \frac{m_{p}}{m_{e}} \frac{\epsilon_{diss}}{c^{2}} \right]^{1/(3-\mu)}\ ,
\label{EqBykov}
\end{equation}
where $\alpha_{M} = 0.1$ -- $1$ is the fraction of the dissipated energy which goes into the magnetic fluctuations; $\zeta$ is the fraction of the electrons which are accelerated and $\mu$ ($1.5 \le \mu \le 2$) is the index of the fluctuation spectrum. For $\zeta \sim 1 $ and $\mu=2$, Eq. (17) corresponds to the usual equipartition assumption, leading to $\Gamma_{e}$ of a few hundreds. In this case, the emission of gamma-rays could 
result from inverse Compton 
scattering on synchotron photons. Bykov and M\'esz\'aros however suggests that only a small fraction
$\zeta \sim 10^{-3}$ of the electrons may be accelerated, leading to $\Gamma_{e}$ values of several thousands.
In this last case, synchrotron radiation can directly produce gamma-rays of typical energy 
\begin{equation}
E_{syn} = 500 \frac{\Gamma_{r}}{300} \frac{B_{eq}}{1000\ {\rm G}} \left(\frac{\Gamma_{e}}{10^{4}}\right)^{2}\ \rm keV\ . 
\end{equation}
\begin{figure}
\resizebox{\hsize}{!}{\includegraphics{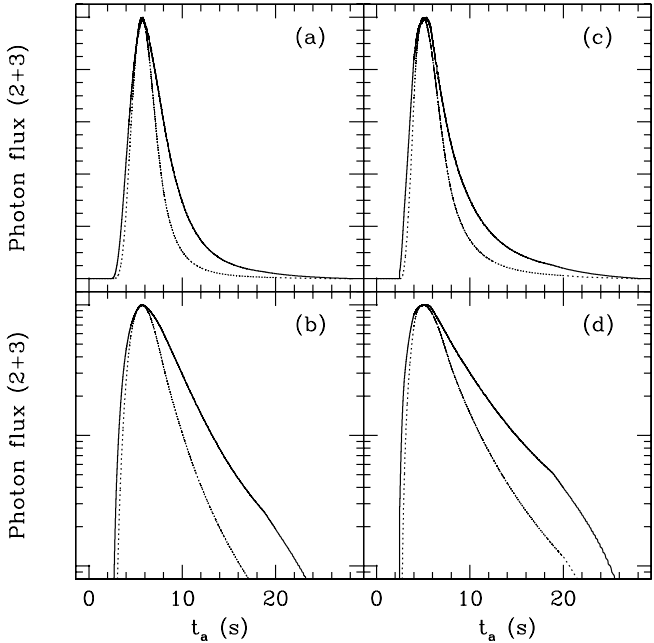}}
\caption{Burst profiles for the initial distribution of the Lorentz factor shown in Fig. \ref{FigInitial}. The photon flux (normalized to the maximum count rate) is given in the interval 50 -- 300 keV, corresponding to BATSE bands 2+3. (a) Profile obtained with the expression of $\Gamma_{e}$ given by Eq. (\ref{EqBykov}); (b) same as (a) in logarithmic scale, which illustrates the exponential decay after maximum; (c) profile obtained with a constant $\Gamma_{e}$; (d) same as (c) in logarithmic scale. In the four panels, the full line represents the profile obtained with 
the hydrocode while the dashed line corresponds to the simple model.}
\label{FigC23_10s}
\end{figure}

We now present a detailed comparison of the results of our hydrodynamical code with those previously obtained 
with the simple model (DM98) for a single pulse burst. We have plotted in
Fig. \ref{FigQuantitesHydro} the values of $t_{e}$, $\Gamma$, $\epsilon_{diss}$ and $\rho$ as function of 
$t_{a}$ for the forward and reverse shocks. We observe an overall similarity 
between the two calculations, 
despite the crude approximations of the simple model. Not surprisingly, the worst estimated quantities 
are the post-shock density and the dissipated energy per proton, which are underestimated by a factor of $\sim 5$. 
Conversely, the emission time and the Lorentz factor of the emitting material are correctly reproduced. 
The emission starts earlier in the simple model where there is no preliminary phase of compression 
before the formation of shocks (this leads to a larger underestimate of the density at the very beginning of the simulation), and ends later.
The total efficiency of the dissipation process is also smaller 
$\sim 5 \%$ instead of $ 12 \%$ for the detailed model.

The other quantities $B_{eq}$, $\Gamma_{e}$ and $e_{syn}$ are not directly given by the hydrodynamical 
simulation but are parametrized by $\alpha_{B}$, $\alpha_{M}$, $\zeta$ and $\mu$, whose values are unknown. To make a useful comparison between the two series of results, we take the same $\alpha_{B}$ and $\mu$ in the two cases but 
adjust $\alpha_{M}/\zeta$ so that the typical synchrotron energy is the same. The corresponding values of $B_{eq}$, $\Gamma_{e}$ and $e_{syn}$ are represented in Fig. \ref{FigEquipartition} with $\alpha_{B}=1/3$, $\mu=1.75$ and $\alpha_{M}/\zeta = 100$ for the hydrocode and $1000$ for the simple model. As expected because of the 
differences in density and dissipated energy, the magnetic field is underestimated by a factor of $5$ in the 
simple model. This is corrected by our choice of
parameters for $\Gamma_{e}$ and the resulting synchrotron energies are very similar in the two cases. Also notice
that the efficiency of
the synchrotron process is smaller in the simple model due to a poor estimate of the mass 
flux accross the shock.

The agreement between the two calculations is satisfactory and allows to
be quite confident in the results of the simple model. 
Compared to the hydrodynamical code, the simple model has very short 
computing times and enables a detailed exploration of the temporal and spectral properties of 
synthetic bursts which was presented in our previous paper (DM98). 
We show in the next section 
the detailed results obtained with the hydrocode in the case of a single 
pulse burst.
\begin{figure}
\resizebox{\hsize}{!}{\includegraphics{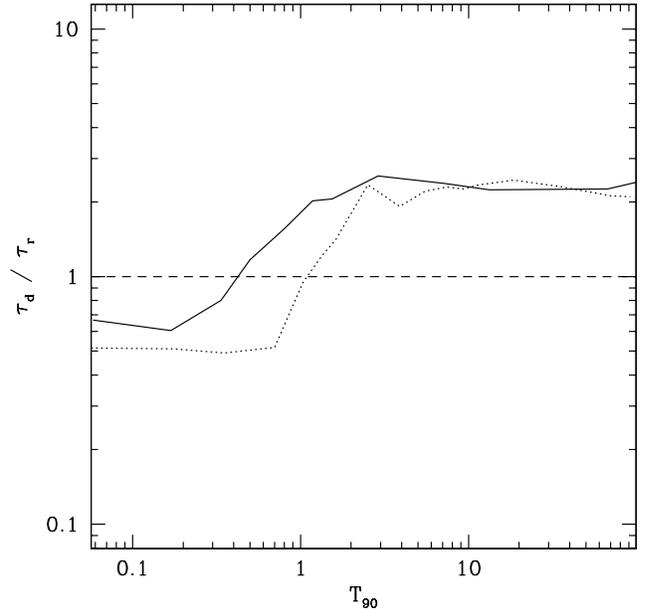}}
\caption{Ratio of the decay to rise times as a function of burst duration. The initial distribution of the Lorentz factor is given by Eq. (\ref{EqGammaInitial}) and the total energy injected into the wind is proportional to $t_{W}$ : $E = \frac{2\ 10^{52}}{4 \pi} \left(\frac{t_{W}}{10\ {\rm s}}\right)\ \rm erg/sr$. The full line corresponds to the results of the hydrocode while the dashed line 
shows the same relation obtained with the simple model.}
\label{FigSymmetry}
\end{figure}

\subsubsection{Temporal properties}
The contributions of the forward and the reverse shocks are added to construct the synthetic burst. We assume that the photons emitted from $t$ to $t+dt$ by an internal shock of current luminosity $L_{sh}$ are distributed according to a simple power-law spectrum
\begin{equation}
\frac{d\left(E n(E)\right)}{dE}\propto \frac{L_{sh}\ dt}{E_{syn}} \left(\frac{E}{E_{syn}}\right)^{-x}\ ,
\label{EqSpectrum}
\end{equation}
where we adopt $x=2/3$ or $x=3/2$ (the two extreme low energy index that are expected for a synchrotron spectrum) for $E<E_{syn}$ and $2<x<3$ for $E>E_{syn}$ ($x=2.5$ in the following). We take into account cosmological effects (time dilation and redshift) assuming that the burst is located at $z=0.5$.

We have plotted in Fig.\ref{FigC23_10s}a the photon flux observed in BATSE bands 2+3 for the initial distribution of 
Lorentz factor shown in Fig.3, 
calculated either with the hydrocode (with $\frac{\alpha_{M}}{\zeta}=100$) or the simple model 
(with $\frac{\alpha_{M}}{\zeta}=1000$). The two profiles look similar but the hydrodynamical code leads to a slower decay. With $t_{5}$ (resp. $t_{95}$) being the time when 5\% (resp 95 \%) 
of the total fluence has been received, we obtain a  duration $T_{90}=t_{95}-t_{5} = 10.4\ \rm s$ instead of $6.67\ \rm s$ with the simple model. Figure \ref{FigC23_10s}b illustrates that the exponential decay of the burst 
is also nicely reproduced with the detailed calculation. However, if we define $t_{max}$ as the time of maximum count rate and $\tau_{r} = t_{max}-t_{5}$ and $\tau_{d} = t_{95}-t_{max}$ as the rise and the decay times, we get 
a ratio $\tau_{d} / \tau_{r} = 2.08$. DM98 found that a larger value of $\tau_{d} / \tau_{r}$ 
and a corresponding profile closer to the characteristic ``fast rise -- exponential decay'' (FRED) shape is obtained by assuming that the fraction $\zeta$ of accelerated electrons increases with the dissipated energy per proton $\epsilon_{diss}$. As in DM98 we adopt $\zeta \propto \epsilon_{diss}$, so that $\Gamma_{e}$ is independent of $\epsilon_{diss}$. Figures \ref{FigC23_10s}c and \ref{FigC23_10s}d
show the resulting profiles with $\Gamma_{e} = 5000$ for the hydrodocode and $\Gamma_{e}=10000$ for the simple model. The profile then better reproduces a typical FRED shape.
\begin{figure}[t]
\resizebox{\hsize}{!}{\includegraphics{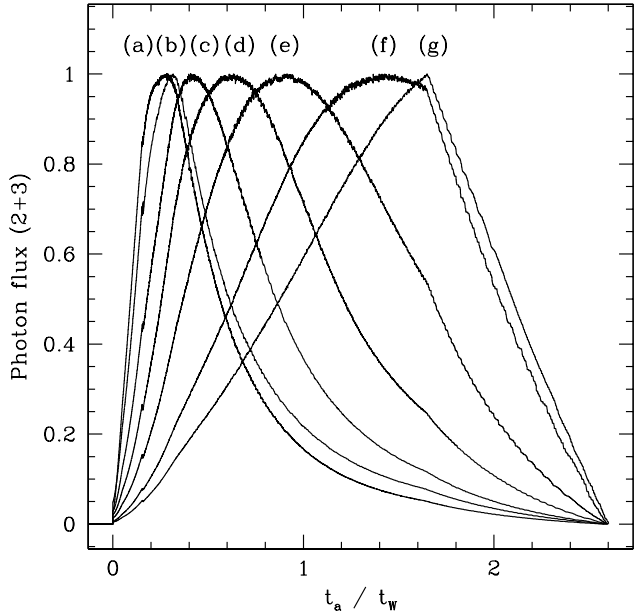}}
\vspace*{-1cm}

\caption{Evolution of the profiles with duration. The initial distribution of the Lorentz factor is given by Eq. (\ref{EqGammaInitial}) and the total injected energy is $E = \frac{2\ 10^{52}}{4 \pi} \left(\frac{t_{W}}{10\ {\rm s}}\right)\ \rm erg/sr$. The different profiles correspond to (a) $t_{W}=10\ \rm s$ ; (b) $t_{W}=5\ \rm s$ ; (c) $t_{W}=2\ \rm s$ ; (d) $t_{W}=1\ \rm s$ ; (e) $t_{W}=0.5\ \rm s$ ; (f) $t_{W}=0.2\ \rm s$ ; (g) $t_{W}=0.1\ \rm s$.}
\label{FigEvolution}
\end{figure}

The observed tendency of short bursts to become symmetric (Norris et al. \cite{Norris1}) has been tested in DM98 with the simple model. The basic behaviour was reproduced but the 
effect was even exagerated since, for $T_{90} < 1\ \rm s$,
$\tau_{d} / \tau_{r}$ was smaller than unity i.e. the decline was faster than
the rise.
As can be seen in Fig. \ref{FigSymmetry} and \ref{FigEvolution}, 
the situation is improved with the hydrocode since now 
$\tau_{d} / \tau_{r} \sim 1$ for $T_{90} \sim 0.4\ \rm s$. However, 
the shortest bursts are still asymmetric with $\tau_{d} / \tau_{r} \sim 0.6$ for $T_{90} \le 0.2\ \rm s$.  

Figure \ref{FigSymmetry} also shows that the ratio $\tau_{d} / \tau_{r}$ is limited to a maximum value of $\sim 2.5$ for the longest bursts
which appears to be in contradiction with the short rise times observed in
some cases. As discussed in DM98, an initial distribution of the Lorentz factor
with a steeper gradient than the one used here  
(Eq. (12)) can indeed increase $\tau_{d} / \tau_{r}$ but
extreme values (such as $\tau_{d} / \tau_{r}$ possibly larger than $10$ in GRB 970208) might still
be difficult to reproduce. 
\begin{figure}
\resizebox{\hsize}{!}{\includegraphics{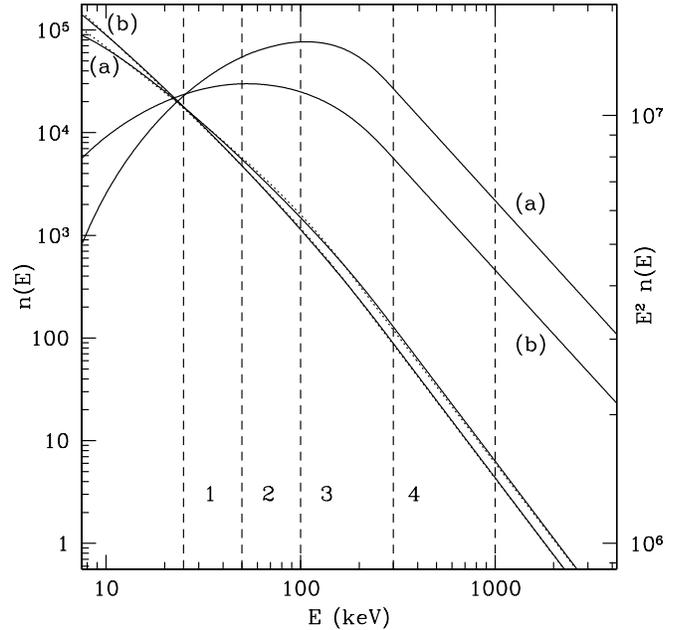}}
\caption{Spectrum of the burst presented in figure \ref{FigEvolution}(c) ($t_{W}=2\ \rm s$). The number of photons per energy interval $n(E)$ and the product $E^{2}\ n(E)$ are shown in arbitrary unit. This product is maximum at peak energy $E_{p} = 403\ \rm keV$ in case (a) ($x=-2/3$) 
and $E_{p} = 193\ \rm keV$ in case (b) ($x=-3/2$). 
The dashed lines show a fit of each spectrum with Band's formula in the interval 10 keV -- 10 MeV (parameters are given in the text).}
\label{FigSpectrum}
\end{figure}
\subsubsection{Spectral properties}
In DM98 we presented a complete study of the global and instantaneous spectral properties of synthetic bursts calculated with the simple model. Since these spectral properties are hardly different when calculated with the hydrocode, we do not present them in detail again. 
We just show in Fig. \ref{FigSpectrum} the shape of the global spectrum calculated for the single pulse burst. 
Despite the very simple form adopted for the instantaneous spectrum (Eq. \ref{EqSpectrum}), the sum of all the 
elementary contributions produces an overall spectrum with a more complex shape, which is well reproduced with Band's formula (Band et al. \cite{Band1})
\begin{eqnarray}
n(E) & = & A \left(\frac{E}{100\ {\rm keV}}\right)^{\alpha} \exp{\left(-\frac{E}{E_{0}}\right)}\, {\rm for}\ E \le (\alpha-\beta) E_{0}\nonumber\\
n(E) & = & A \left[\frac{(\alpha-\beta) E_{0}}{100\ {\rm keV}}\right]^{\alpha-\beta} \exp{\left(\alpha-\beta\right)} \left(\frac{E}{100\ {\rm keV}}\right)^{\beta}\,\nonumber\\
& & {\rm for}\ E \ge (\alpha-\beta) E_{0}\ .
\end{eqnarray}
We find values of the parameters comparable to those observed in real bursts. The best fits in 
Fig. \ref{FigSpectrum} correspond 
to $\alpha = -0.935$, $\beta = -2.42$ and $E_{0} = 239\ \rm keV$ in case (a) ($x=-2/3$) and $\alpha = -1.60$, $\beta = -2.47$ and $E_{0} = 609\ \rm keV$ in case (b) ($x=-3/2$).
\begin{figure*}
\resizebox{8cm}{!}{\includegraphics{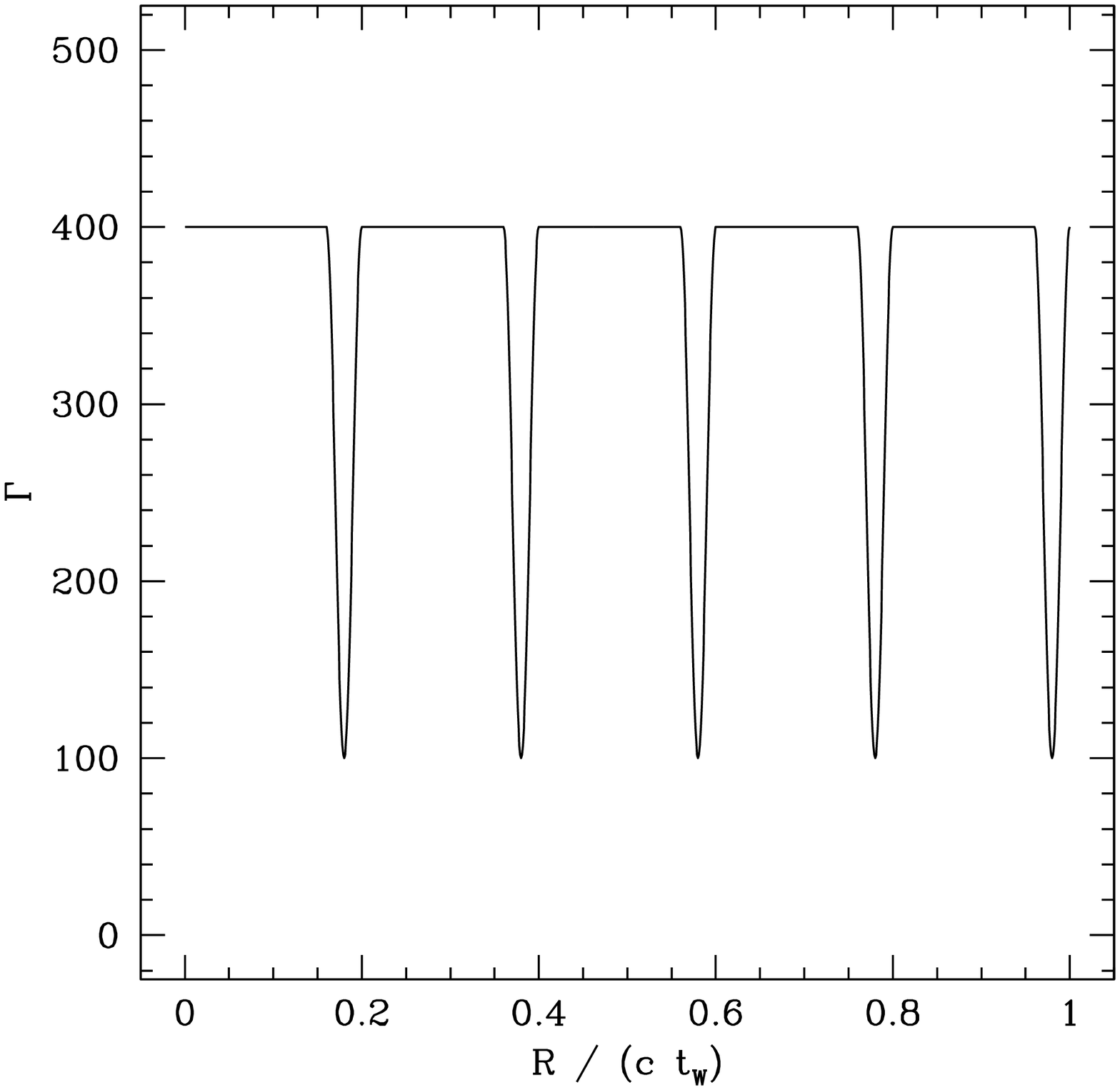}}
\hspace*{1.5cm}
\resizebox{8cm}{!}{\includegraphics{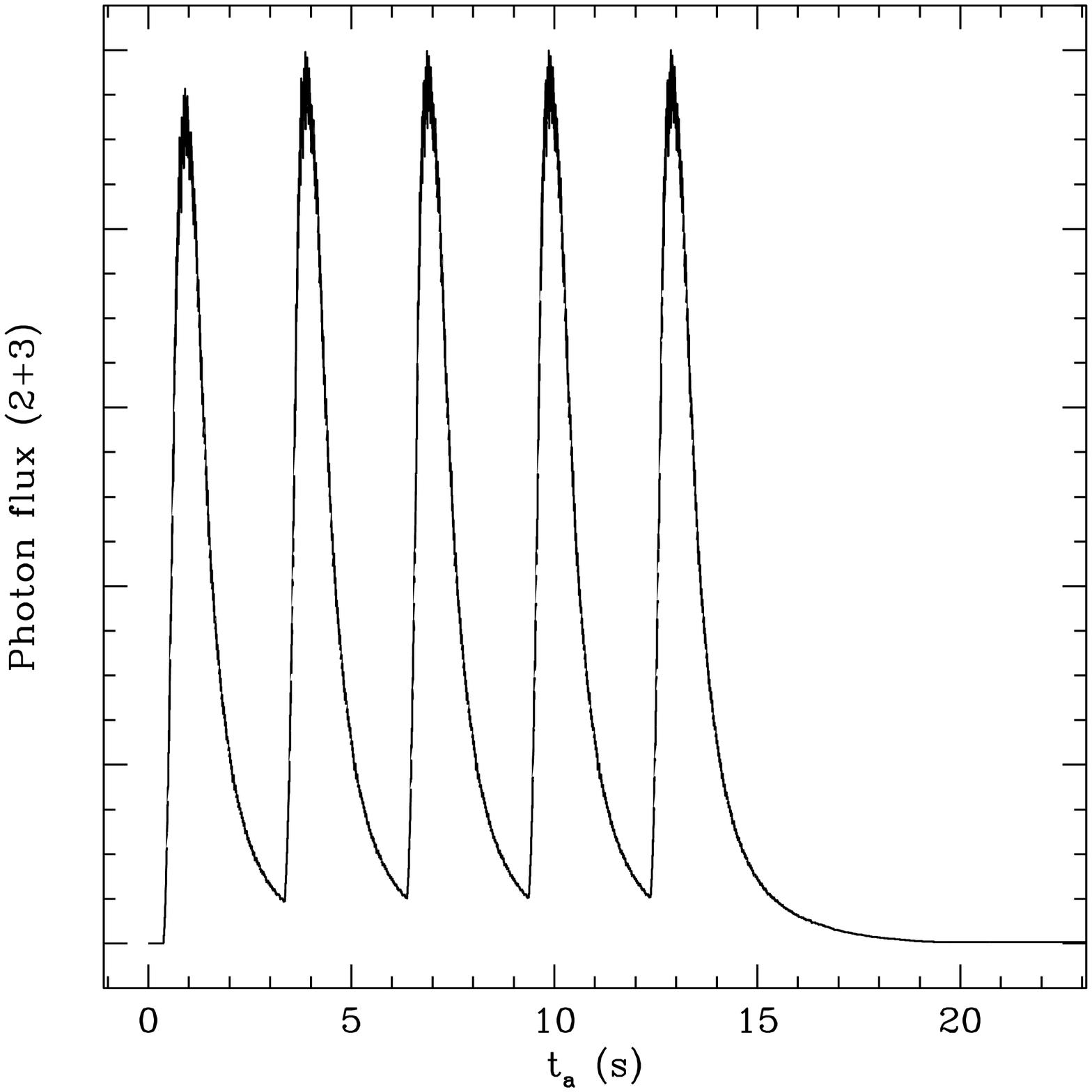}}
\caption{Example of a complex burst. Left panel: initial distribution of the 
Lorentz factor with five identical patterns. Right panel: corresponding profile obtained with the hydrocode (normalized photon flux in the interval $50$ - $300\ \rm keV$). The wind is produced with a duration $t_{W}=10\ \rm s$ and the injected energy is $E=10^{52}/4\pi\ \rm erg/sr$.}
\label{FigComplex1}
\end{figure*}
  As $x$ is limited to the range $2/3 < x < 3/2$, we cannot get
  spectra with low energy slopes flatter than $-2/3$ as they are 
  observed in several bursts (Preece et al. \cite{Preece1}). A more detailed
  description of the radiative processes is then needed to reproduce 
  these extreme slopes (an attempt to solve this problem is proposed by
  Meszaros \& Rees \cite{Meszaros6}).\\
  However, even with the crude modelization of the instantaneous spectra 
  which is used here, it has been shown in DM98 that several spectral 
  properties of GRBs are reproduced. In particular, the hard to soft 
  evolution during a pulse and the change of pulse shape as a function of 
  energy as well as the duration -- hardness ratio relation which appears 
  as a natural consequence of the internal shock model. These important 
  spectral features are confirmed in our detailed hydrodynamical calculation.

\subsection{Case of more complex bursts}
An important property of the internal shock model is its ability to produce a great variety of temporal profiles. 
Norris et al. (\cite{Norris1}) have shown that complex bursts can generally be analysed in terms of a series of (possibly overlapping) simple pulses. 
This result is readily interpreted in the context of the internal shock model. A wind made of a succession of
fast and slow shells will produce a succession of pulses which will add to form a complex burst. 

We present such examples of complex bursts in Fig. 11 and 12. The first one (Fig. \ref{FigComplex1}) is produced by an initial distribution of the Lorentz factor made of five consecutive identical patterns. Each pattern made of a slow and 
a
rapid part produces its own individual pulse and the resulting burst 
 has a complex shape with five, very similar, pulses. Our second example 
(Fig. 12) uses the same type of initial distribution of the Lorentz factor 
but the slow parts now have non equal $\Gamma$ values. The resulting burst 
is more realistic with four pulses of different intensities.

We did not treat with the hydrocode a large number of cases as we did with 
the simple model. Nevertheless we confirmed the essential result that the
variability introduced in the initial distribution of the Lorentz factor
is present in the burst profile with the same time scale. The profile therefore 
appears as a direct indicator of the activity of the central engine.
\begin{figure*}
\resizebox{8cm}{!}{\includegraphics{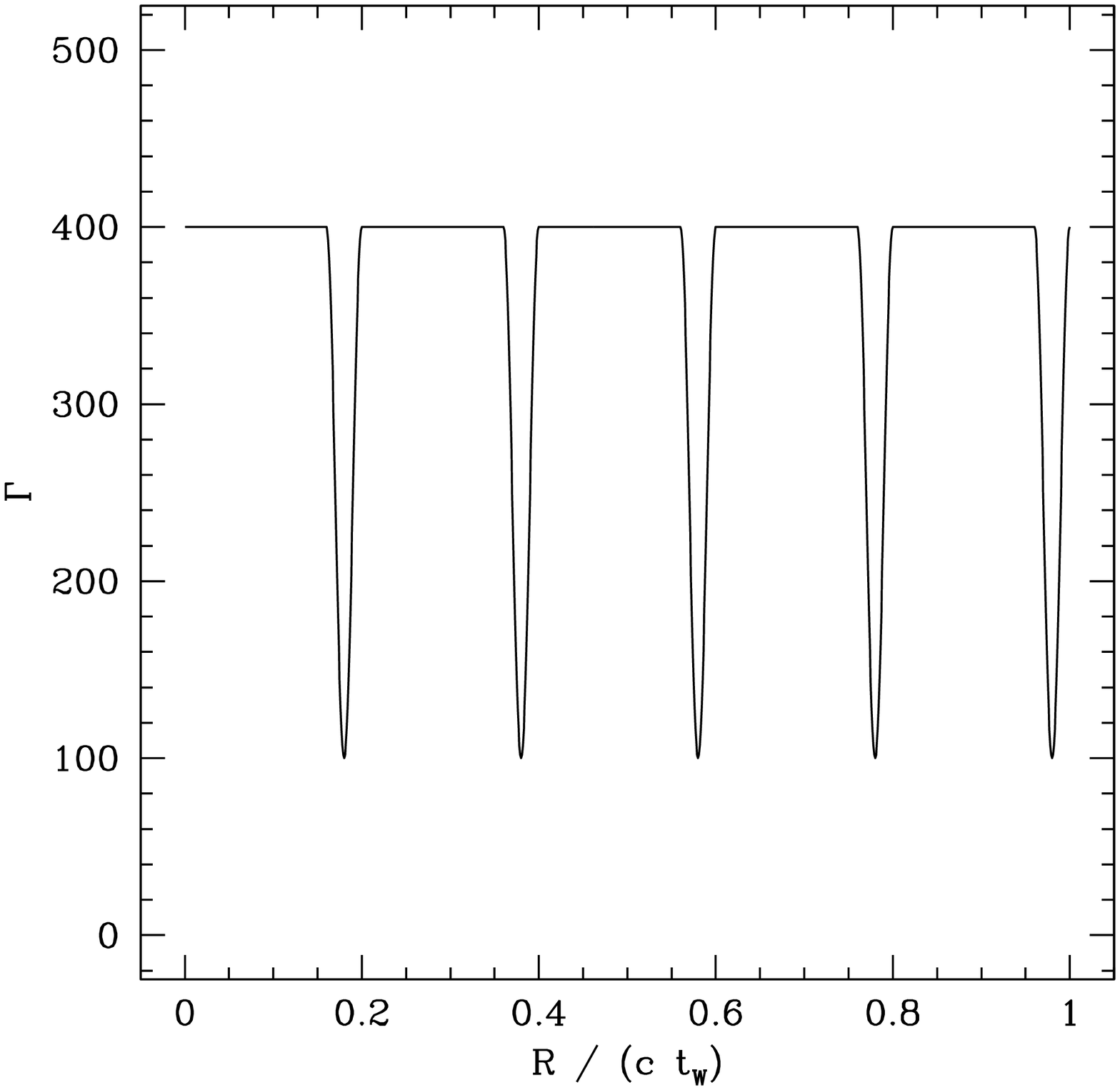}}
\hspace*{1.5cm}
\resizebox{8cm}{!}{\includegraphics{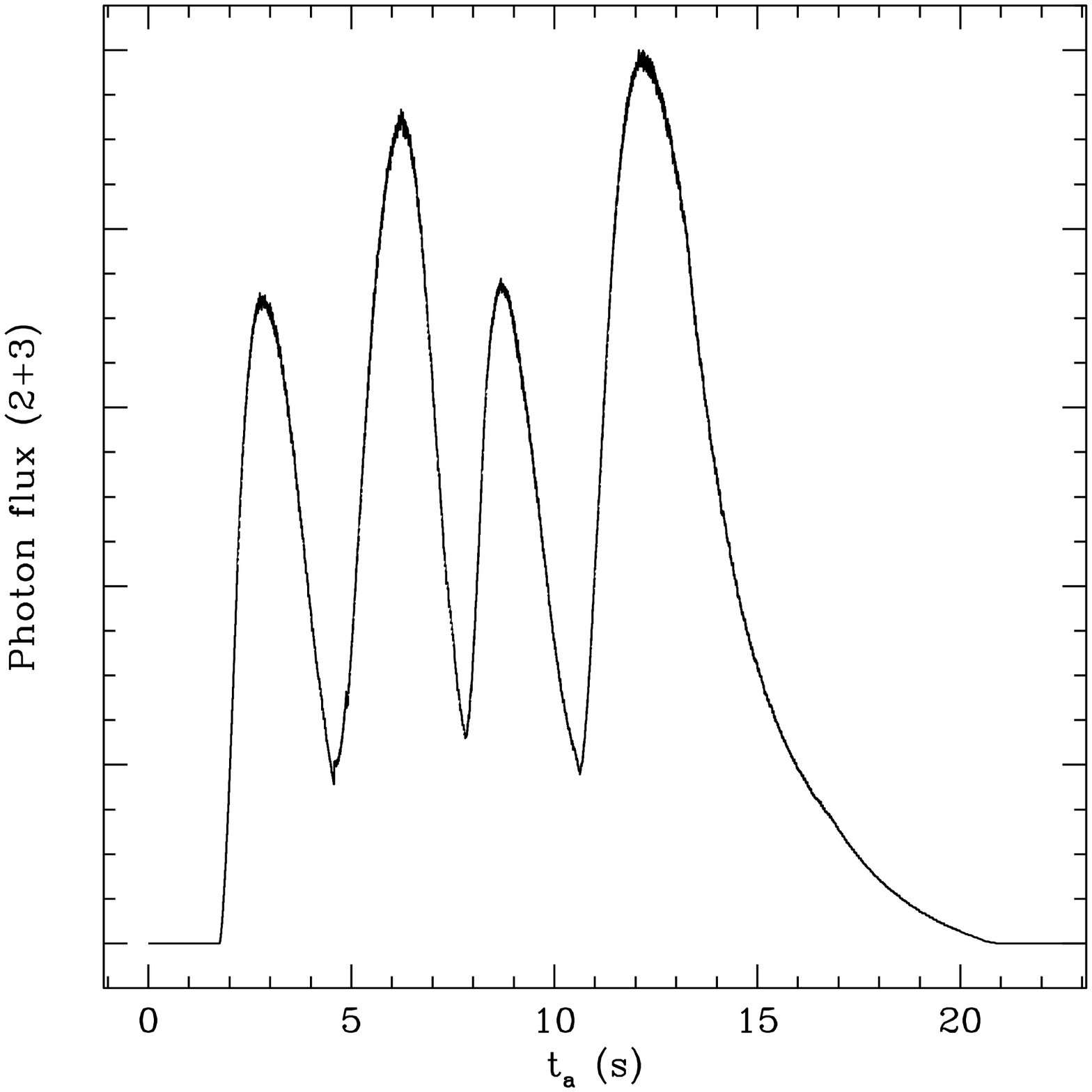}}
\caption{Another complex burst. Left panel: initial distribution of the Lorentz factor with now four non identical patterns (Lorentz factors in the
slow parts are different). Right panel: corresponding profile obtained with the hydrocode (normalized photon flux in the interval $50$ - $300\ \rm keV$). The wind is produced with a duration $t_{W}=10\ \rm s$ and the injected energy is $E=10^{52}/4\pi\ \rm erg/sr$.}
\label{FigComplex2}
\end{figure*}

\section{Conclusions}
This paper is the continuation of our study of the internal shock model started in DM98. We developed a 1D lagrangian relativistic hydrocode (in spherical symmetry) to validate our previous simpler approach where all pressure
waves were neglected in the wind. Our code is an extension of the classical \textit{PPM} method of Colella and Woodwards (\cite{Colella1}) in the spirit 
of the work by Mart\'{\i} \& M\"uller (\cite{Marti2}) for the eulerian case in planar symmetry. 

A detailed comparison has been made between the hydrocode and the simple model 
in the case of a single pulse burst. It appears that the dynamical evolution of the wind is well reproduced by the simple model, which is not too surprising 
because the wind energy is largely dominated by the kinetic part so that the effect of pressure waves is small. Only
one physical quantity -- the density of the shocked material -- is strongly underestimated in the simple model. 
In order to make valuable comparisons between the two calculations we 
have therefore
adjusted the equipartition parameters so that the mean value of the synchrotron energy is the same in the two cases. The synthetic bursts which are then 
obtained are very similar which proves that our first approach was 
essentially correct and confirm our previous results. A similar conclusion
was reached by Panaitescu and M\'esz\'aros (\cite{Panaitescu3}) who performed a
comparable study.

The internal shock model can easily explain the great temporal variability observed 
in GRBs.  
The main characteristic features of individual pulses are well reproduced: 
(1) pulses have typical asymmetric ``FRED'' profiles; (2) the pulse width 
decreases with energy following a power--law $W(E)\propto E^{-p}$ with $p \sim 0.4$; (3) short pulses show a tendency to become more symmetric. Our model still 
gives very short pulses which decay faster than they rise but the hydrodynamical simulation improves the situation compared to the simple model. 
Spectral properties of GRBs are also well reproduced. We obtain synthetic 
spectra which can be nicely fitted with Band's function with parameters 
comparable to those observed in real GRBs. The spectral hardness and the count rate are correlated during the evolution of a burst with the hardness usually 
preceeding the count rate. As also pointed in DM98, the duration--hardness relation is a natural consequence of the internal shock model.
These results are very encouraging and the main difficulty which remains is the low efficiency (about $10 \%$) of the internal shock model. As long as the energetics of GRBs and the mechanism initially operating in the central engine 
are not precisely identified, we cannot say if this is a critical problem or not.
We still believe that the internal shock model is at present the most convicing candidate to explain the gamma--ray emission from GRBs.

Next steps in this work will address the following questions. We first want to extend our hydrodynamical code to a non--adiabatic version in order to include the 
radiative losses in the dynamical calculation. We have already developed an 
``isothermal Rieman Solver'' for that purpose (Daigne \& Mochkovitch \cite{Daigne1}). We would also like to 
study the effects of the external medium, with a special attention to the 
reverse shock which propagates into 
the wind and possibly interacts with the internal shocks. 
Preliminary results with the simple method using ``solid layers'' have already been 
obtained (Daigne \& Mochkovitch \cite{Daigne4}) but they have to be confirmed by a hydrodynamical 
calculation. 
Finally, we would like to investigate the details of the emission process during internal shocks to solve some of the problems encountered by the synchrotron model.


\begin{acknowledgements}
We would like to thank the Departamento de Astronom\'{\i}a y Astrof\'{\i}sica de Valencia for their kind hospitality and acknowledge clarifying discussions with J.M. Mart\'{\i} and J.M. Ib\'a\~nez.
\end{acknowledgements}

\end{document}